\documentclass[pra, showpacs,twocolumn, superscriptaddress,groupedaddress, noshowpacs]{revtex4}  % for review and submission

\usepackage{hyperref}
\usepackage{amsmath,amssymb,amsfonts,amsthm}
\usepackage{braket}
\usepackage{cancel}
\usepackage{color}
\usepackage{graphicx}
\usepackage{comment}

\usepackage{physics}

\renewcommand{\vec}[1]{\ensuremath{\boldsymbol #1 }}

\newcommand {\fexp} [1] {\exp \left( #1 \right)}

\begin{document}

\title{Twist-and-turn dynamics of spin squeezing in bosonic Josephson junctions: Enhanced shortcuts-to-adiabaticity approach}

\author{Manuel Odelli}
\affiliation{School of Physics, University College Cork, Cork, Munster T12 K8AF, Ireland}

\author{Andreas Ruschhaupt}
\address{School of Physics, University College Cork, Cork, Munster T12 K8AF, Ireland}

\author{Vladimir M. Stojanovi\'c}
\address{Institut f\"{u}r Angewandte Physik, Technical University of Darmstadt,
64289 Darmstadt, Germany}

\begin{abstract}
The twist-and-turn dynamics of spin squeezing results from the interplay of the one-axis-twisting (nonlinear
in the collective-spin operators) and the transverse-field turning (linear) term in the underlying Lipkin-Meshkov-
Glick-type Hamiltonian, both with constant (time-independent) prefactors. Using shortcuts to adiabaticity (STA)
and the recently developed enhanced version thereof (eSTA), we demonstrate here that dynamics of this type
can be utilized for a fast and robust preparation of spin-squeezed states in internal bosonic Josephson junctions,
i.e., condensates of cold bosonic atoms in two different internal (hyperfine) states (single-boson modes) coupled
through Rabi rotations. Assuming that the initial state of this system is its ground state for a given initial value
of the (time-dependent) linear coupling strength and that the nonlinear coupling strength remains constant, we
set out to determine the time dependence of the linear (Rabi) coupling strength that allows for the generation of
spin-squeezed states using the STA- and eSTA-based approaches. We then characterize the modified twist-and-
turn dynamics of this system by evaluating the coherent spin-squeezing and number-squeezing parameters, as
well as the fidelity of the target spin-squeezed states. In this way, we show that the eSTA approach allows for
a particularly robust realization of strongly spin-squeezed states in this system, consistently outperforming its
adiabatic and STA-based counterparts, even for systems with several hundred particles. Our method could also
be employed for the generation of metrologically-useful non-Gaussian states.
\end{abstract}

\maketitle

\section{Introduction} 
Recent years have witnessed tantalizing progress in the realm of quantum-state engineering~\cite{StojanovicPRL:20,Peng+:21,StojanovicPRA:21,
Zheng++:22,Haase++:22,Zhang+:23,Zhang++:23,Erhard+:18,Haase+:21,Qiao+:22,Feng+:22,Nauth+Stojanovic:22,
Stojanovic+Nauth:22, Stojanovic+Salom:19,Stefanatos+Paspalakis:20, 
Ying:24,Wu+:17,Stojanovic+Nauth:23,Saleem+:24,Chen+:24}. Being highly
interwoven with the constant improvement of methods for manipulation and 
control of quantum systems, this steady progress remains of pivotal importance for further development of quantum-enhanced technologies 
in various physical platforms~\cite{Dowling+Milburn:03}. In particular, prompted by the anticipated quantum-technology applications of highly 
entangled multiqubit states a large variety of schemes for fast (compared to the relevant coherence time of the underlying physical 
system) and robust (against various system-specific sources of decoherence 
and noise) generation of such states have been proposed
in recent years, especially for $W$~\cite{StojanovicPRL:20,Peng+:21,StojanovicPRA:21,Zheng++:22,Haase++:22,Zhang+:23,Zhang++:23}, 
Greenberger-Horne-Zeilinger (GHZ)~\cite{Erhard+:18,Haase+:21,Qiao+:22,Feng+:22,Nauth+Stojanovic:22,Stojanovic+Nauth:22}, and Dicke~\cite{Wu+:17,Stojanovic+Nauth:23,Saleem+:24} states. Another important example of highly-entangled quantum many-body states of interest for quantum-enhanced metrology is furnished by spin-squeezed states~\cite{Kitagawa+Ueda:93}.

% ------- Spin-squeezed states -------------
Proposed in the seminal work of Kitagawa and Ueda~\cite{Kitagawa+Ueda:93}, the concept of spin squeezing -- i.e., that of 
redistributing the spin fluctuations between two orthogonal directions -- was demonstrated to allow an enhancement in the precision of 
atom interferometers~\cite{Wineland+:92,Wineland+:94}. Much like their photonic counterparts in optical interferometry~\cite{Caves:81}, 
spin-squeezed states provide phase sensitivities beyond the standard quantum limit (SQL) for phase uncertainty 
$\Delta\Theta_{\textrm{SQL}}=1/\sqrt{N}$, the latter being characteristic of probes involving a finite number $N$ of uncorrelated 
(or classically correlated) particles~\cite{Giovanetti+Lloyd+Maccone:06}. Owing to the fact that they allow one to overcome 
this classical bound, spin-squeezed states established themselves as a valuable resource for quantum metrology~\cite{Pezze+:18}. 
Subsequently, the important link between spin squeezing and entanglement~\cite{Li:22} -- the ingredient that spin-squeezed states share with other types of states defying the SQL (for instance, GHZ states~\cite{Lee+Kok+Dowling:02,Saleem+:24}) -- was also established~\cite{Sorensen+:01,Sorensen+Moelmer:01}.

In their original work, Kitagawa and Ueda investigated the preparation of spin-squeezed states based on 
one-axis twisting (OAT) and two-axis-countertwisting (2ACT) Hamiltonians, where ``twisting'' refers to terms 
that are quadratic in the collective-spin operators~\cite{Kitagawa+Ueda:93}. While both of these Hamiltonians 
permit the generation of spin-squeezed states, they show qualitatively different behavior. The OAT Hamiltonian 
does not saturate the fundamental quantum-metrological bound of sensitivity and leads to the maximal squeezing 
at a time that scales as $J^{1/\mu}$, where $J$ is the size of the collective spin and $\mu>0$. In contrast to 
its OAT counterpart, the 2ACT Hamiltonian does saturate the fundamental quantum-mechanical limit on sensitivity 
and permits the generation of maximal spin squeezing in a time that is logarithmic in the collective-spin size. 
Subsequently, a model that in addition to the (nonlinear) OAT term involves a linear term describing a transverse 
field was also investigated by Micheli {\em et al.}~\cite{Micheli+:03}. For this type of spin-squeezing Hamiltonian, which gives rise to what became known as twist-and-turn (TNT) dynamics~\cite{Haine+Hope:20,Huang+:22}, they demonstrated that it allows the preparation of highly entangled metrologically-relevant states (cat-like states)~\cite{Strobel+:14} at times that 
are -- similar to the 2ACT case -- logarithmic in the size of the 
collective spin~\cite{Micheli+:03}. Finally, in a study complementary 
to that of Ref.~\cite{Micheli+:03}, it was demonstrated that this type of dynamics is optimal as far as the 
generation of spin squeezing is concerned~\cite{Sorelli+:19}, at least in the absence of decoherence and losses.

The most natural physical setting in which to demonstrate the TNT dynamics of spin squeezing 
and harness the latter for generating strongly spin-squeezed states is provided by Bose-Einstein condensates 
(BECs) of cold neutral atoms. Indeed, the generation of spin-squeezed states was demonstrated more than a decade ago
in proof-of-principle experiments with interacting cold $^{87}$Rb atoms in bosonic Josephson junctions (BJJs) 
~\cite{Esteve+:08,Gross+:10,Riedel+:10,Zibold+:10}; these systems, where bosons within a condensate can be restricted 
to occupy only two single-atom states (modes), come in two varieties: internal BJJs [where the two relevant modes 
correspond to two different internal (hyperfine) atomic states, with a linear, Rabi-type coupling between them] 
and external ones (in which the two modes correspond to bosons trapped in two spatially separated wells of an 
external double-well potential)~\cite{Gati+Oberthaler:07}. The TNT-type dynamics was experimentally
investigated in an internal BJJ, assuming a constant linear-coupling strength and an abrupt change (i.e. a quench)
of a nonlinear-coupling strength from zero to a finite value~\cite{Muessel+:15}. Finally, an alternative approach
for generating spin squeezing in internal BJJs has quite recently been theoretically proposed~\cite{Odelli+:23}, 
which makes use of the methods of shortcuts to adiabaticity (STA)~\cite{Chen2010STA,chen2010b,ibanez2012,torrontegui2013a, 
STA_RMP:19} and their recently proposed enhanced version (eSTA)~\cite{Whitty+:20,Whitty+:22, Whitty2022b}.

% ----- Review eSTA ---------------

STA are a family of analytical control techniques that mimic adiabatic evolutions, but are typically much faster 
than their adiabatic counterparts~\cite{Chen2010STA,chen2010b,ibanez2012,torrontegui2013a, STA_RMP:19}. Generally speaking, 
analytical control approaches are highly desirable because they are usually rather
simple, in addition to providing greater physical insight and allowing for a superior stability under various 
experimental imperfections~\cite{ruschhaupt2012a,lu2020}. STA have heretofore been utilized in many different contexts~\cite{li2022,kiely2016,kiely2018,torrontegui2011,hasan2024}, including 
that of engineering spin-squeezed states in BJJs~\cite{JuliaDiaz++:12}; besides, their use was also proposed for engineering other types 
of entangled states (such as, e.g., NOON states~\cite{Hatomura:18,Stefanatos+Paspalakis:18}). They proved to yield 
results comparable to those obtained using optimal-control methods~\cite{Lapert+Ferrini+Sugny:12,Sorelli+:19}. 

One typical limitation of STA methods is that they often require non-trivial physical implementation (e.g. counterdiabatic driving~\cite{STA_RMP:19});
likewise, some STA techniques can only be straightforwardly applied to small or highly symmetric systems (e.g. methods based on 
Lewis-Riesenfeld 
invariants)~\cite{STA_RMP:19}. This served as the primary motivation behind the development of Enhanced Shortcuts to Adiabaticity (eSTA) \cite{Whitty+:20, Whitty+:22, Whitty2022b}, an approach that 
allows one to perturbatively improve STA solutions in an analytical fashion and, even more importantly, 
design efficient control protocols for systems in which STA are not directly applicable. The eSTA method was shown to outperform its STA counterparts in nontrivial quantum-control problems related to coherent atom transport in optical
lattices~\cite{Whitty+:20,Hauck+:21,Hauck+Stojanovic:22} and anharmonic trap expansion~\cite{Whitty2022b}. Besides, it was demonstrated that eSTA-based control schemes are more robust against various types of imperfections than those based on STA methods~\cite{Whitty+:22}.

In this paper, we revisit the problem of efficiently preparing spin-squeezed states in internal BJJs by modifying the underlying TNT-type dynamics of spin squeezing in this system.
We also assume that the 
nonlinear-coupling strength in this system remains constant (i.e. time-independent) and subsequently determine the time-dependence of the 
linear-coupling strength that allows the preparation of spin-squeezed states using the STA and eSTA methods; we also compare the performance of the latter methods with their adiabatic counterpart - a simple linear sweep of the linear-coupling strength from a large initial- to a small final value. We then quantify the state-preparation process for the desired spin-squeezed states in this system by computing the values of the coherent spin-squeezing- and number-squeezing parameters, as well as the target-state fidelity. In this way, we show that our proposed eSTA-based control scheme allows for a particularly robust realization of strongly spin-squeezed states in internal BJJs. Importantly, we demonstrate that this approach consistently outperforms its STA-based and adiabatic counterparts, even for particle numbers that are in the range of several hundreds.  

The remainder of this paper is organized in the following manner. In Sec.~\ref{SystemProblem}, we set the stage for 
further discussion by briefly reviewing the essential physics of internal BJJs and introducing their underlying 
Lipkin-Meshkov-Glick-type Hamiltonian; we also introduce the relevant figures of merit for characterizing spin squeezing. 
Section~\ref{ApplicationESTA} is dedicated to the description of three different control schemes that we employ for the generation of spin-squeezed states in the present work:
an established STA-based scheme, its improved version, and,
finally, an eSTA-based scheme. In Sec.~\ref{sect_results} we present the results for the target-state fidelities and spin-squeezing parameters obtained within the proposed eSTA-based scheme; we also compare the obtained results with those corresponding to the conventional- and improved STA approaches, 
as well as the simplistic adiabatic-sweep approach. In this section, we also demonstrate the superior robustness of our proposed, eSTA state-preparation scheme to the STA-based one. Finally, we also compare the modified TNT-type dynamics of
spin squeezing inherent to our approach with the well-established OAT dynamics. The paper is summarized, 
with some concluding remarks and outlook, in Sec.~\ref{SummConcl}. Some mathematical details, pertaining to the derivation of 
eigenfunctions of STA invariants in momentum space and their Fourier transforms, are relegated to Appendix~\ref{AppendMomSpace}.

% ---- Fig. 1: Sketch plot of the two hyperfine states in the internal BJJ ----
\begin{figure}[b!]
\centering
\includegraphics[width=0.8\linewidth]{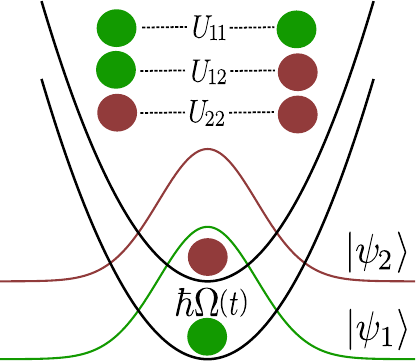}
\caption{\label{fig:sketch}Pictorial illustration of an internal BJJ, i.e. a condensate of bosons in two 
different hyperfine states (modes) $|\psi_1\rangle$ and $|\psi_2\rangle$. An external electromagnetic field 
coherently transfers atoms between these two modes (Rabi rotations); the corresponding linear (Rabi-type) 
coupling strength $\hbar\Omega$ is assumed to depend on time. The intraspecies interaction strengths $U_{11}$ 
and $U_{22}$, as well as their interspecies counterpart $U_{12}$, are assumed to be time-independent. The value 
of $U_{12}$ can be suppressed by means of an external magnetic field in the presence of a Feshbach resonance, 
leading to an increase in the nonlinear coupling strength $\chi=(U_{11}+U_{22} -2U_{12})/\hbar$.} 
\end{figure}

\section{System and spin-squeezed states} \label{SystemProblem}
To set the stage for further discussion, in what follows we first provide a short introduction into the system under consideration -- internal BJJs -- and briefly review the essential concepts pertaining to spin squeezing (Sec.~\ref{IntroIBJJ}). We then concisely introduce two of the most relevant figures of merit of spin squeezing -- the coherent-spin-squeezing- and number-squeezing parameters -- and summarize their basic properties (Sec.~\ref{SpinSqueezingParams}).

\subsection{Internal BJJs and twist-and-turn Hamiltonian} \label{IntroIBJJ}
An internal BJJ is created with a trapped BEC that consists of cold atoms in two different internal (hyperfine) 
states~\cite{Steel+Collett:98} (for a pictorial illustration, see Fig.~\ref{fig:sketch} below). A typical example 
of such a system is a condensate of $^{87}$Rb atoms, where the role of the two relevant internal states can, for 
example, be played by the $|F=1, m_{\textrm F}=1\rangle$ and $|F=2, m_{\textrm F}=-1\rangle$ hyperfine sublevels 
of the electronic ground state of rubidium. Assuming that the external atomic motion in such a system is not 
affected by internal dynamics, it is pertinent to use a single-mode approximation for atoms in each of the two 
hyperfine states; these two single-boson modes will in the following be denoted by $|\psi_1\rangle$ and $|\psi_2\rangle$. An internal BJJ is typically prepared by trapping the atoms in the two internal states
within the wells of a deep one-dimensional optical lattice; the depth of such a lattice ought to be sufficiently 
large that coherent tunnelling of atoms between different wells is suppressed~\cite{Gross+:10}. 

The linear, Rabi-type, coupling of atoms in the two relevant internal states of an internal BJJ is enabled 
by an electromagnetic field that coherently transfers atoms between those states by means of Rabi rotations~\cite{Hall+:98}. 
At the same time, atoms interact through $s$-wave two-body interactions (nonlinear coupling), both atoms in the
same internal state (intraspecies interaction) and those in different states (interspecies interaction). As a result, an internal BJJ is described by a two-state Bose-Hubbard model~\cite{Stojanovic+:08,Hofer+:12}.
Using the Schwinger-boson formalism (see, e.g., \cite{Pezze+:18}), the corresponding Hamiltonian can be written in the form
\begin{equation}\label{eq:HamiltonianIBJJ}
H_{\textrm{IBJJ}}(t) =  \hbar \chi  J_z^2 - \hbar \Omega (t)  J_x \:,
\end{equation}
which represents a special case of the Lipkin-Meshkov-Glick family of Hamiltonians~\cite{Lipkin+Meshkov+Glick:65}.
Here $\hbar\Omega$ is the strength of Rabi-type coupling; in the problem at hand, this linear-coupling strength
is assumed to depend on time, i.e., $\Omega=\Omega(t)$. The nonlinear-coupling strength $\chi$ is assumed to be constant and positive (corresponding to repulsive two-body interactions between atoms) in the following. 
$ \mathbf{J}\equiv\{ J_x, J_y, J_z\}$ are pseudoangular-momentum operators describing
the collective spin of bosons, which is defined as $ \mathbf{J} =\sum_{k} \vec{\sigma}^{(k)}$, 
with $ \vec{\sigma}^{(k)}\equiv \{ \sigma^{(k)}_x, \sigma^{(k)}_y, \sigma^{(k)}_z\}$
being the Pauli operators representing the pseudospin degree of freedom of the $k$-th atom ($k=1,2,\ldots, N$). 
These pseudoangular-momentum operators, which satisfy the standard commutation relation $[J_{\alpha},J_{\beta}]=i\hbar\sum_{\gamma} \epsilon_{\alpha\beta\gamma} J_\gamma$ of the $su(2)$ algebra (where $\alpha,\beta,\gamma=x,y,z$ and            $\epsilon_{\alpha\beta\gamma}$ is the Levi-Civita symbol), 
are given by
\begin{eqnarray} 
J_x &=& \frac{1}{2}\:(J_{+} + J_{-}) \:, \nonumber \\
J_y &=& \frac{1}{2i}\:(J_{+} -  J_{-}) \:, \label{PseudoAngular}\\
J_z &=& \frac{1}{2}\:(n_1 - n_2) \nonumber \:.
\end{eqnarray}
In terms of the creation and annihilation operators $ a^{\dagger}_i$ and $ a_i$ ($i=1,2$)
corresponding to the two relevant single-boson modes, the ladder operators $ J_{\pm}$ are expressed
as $ J_{+}= a^{\dagger}_2  a_1$, $ J_{-}= a^{\dagger}_1  a_2$, while $ n_1= a^{\dagger}_1  a_1$ and $ n_2= a^{\dagger}_2 
 a_2$ are the particle-number operators. 

It is important to point out that in the ground state of a system described by the Hamiltonian with 
$J_z^2$ and $J_x$ terms [cf. Eq.~\eqref{eq:HamiltonianIBJJ}], the mean collective spin points in the 
direction of $J_x$ and the direction of minimal variance is that of $J_z$ [cf. Eq.~\eqref{PseudoAngular}]. 
While the maximal length of the mean collective spin -- achieved in the complete absence of nonlinear 
coupling -- is $N/2$, in the following we will be discussing cases with finite nonlinear-coupling strength, 
including those in which the mean collective-spin length is significantly reduced compared to this maximal 
value.

The linear-coupling strength in the Hamiltonian $H_{\textrm{IBJJ}}(t)$ [cf. Eq.~\eqref{eq:HamiltonianIBJJ}]
is given by~\cite{Pezze+:18}
\begin{equation} \label{LinearCouplParameter}
\Omega =  \Omega_R \int d^{3}\mathbf{r}\:\psi^{*}_1(\mathbf{r})\psi_2(\mathbf{r}) \:, 
\end{equation}
where $\Omega_R$ stands for the Rabi frequency and $\psi_{i}(\mathbf{r})\equiv \langle
\mathbf{r}|\psi_{i}\rangle$ ($i = 1, 2$) are the two internal states in the coordinate 
representation (i.e. the mode functions of those states). The time dependence of the linear-coupling strength $\Omega=\Omega(t)$ can be manipulated with a high degree of control; for instance, this can be accomplished experimentally by controlling the magnitude of the electromagnetic field used. Importantly, the existing experimental capabilities allow for making rapid changes in both the amplitude and the phase of $\Omega(t)$~\cite{Muessel+:15}.

The nonlinear-coupling strength $\chi$ is given 
by~\cite{Pezze+:18,Hofer+:12}
\begin{equation} \label{NonlinearCouplParameter}
\hbar\chi = U_{11}+ U_{22} - 2 U_{12}\:, 
\end{equation}
where $U_{ij}$ ($i,j = 1, 2$) are the (two-body) $s$-wave intraspecies ($i=j$) and 
interspecies ($i\neq j$) interaction strengths [cf. Fig.~\ref{fig:sketch}]. The latter can be expressed as 
\begin{equation} \label{InteractionStrengths}
U_{ij} = \frac{2\pi\hbar^2 a^{(i,j)}_s}{M} \int d^{3}\mathbf{r}|\psi_i(\mathbf{r})|^2
|\psi_j(\mathbf{r})|^2 \:,
\end{equation}
where $a^{(i,j)}_s$ ($i,j = 1, 2$) are the corresponding $s$-wave scattering lengths 
and $M$ is the mass of a single atom. It is important to stress 
that the interspecies $s$-wave scattering length $a^{(1,2)}_s$ for $^{87}$Rb atoms can be tuned using 
an external magnetic field owing to the presence of Feshbach resonance~\cite{Zibold+:10}; this mechanism can be utilize to reduce the interspecies $s$-wave scattering length, given that for $^{87}$Rb atoms 
there is a nearly perfect compensation of intraspecies and interspecies interactions. 
Another approach for tuning the nonlinear-coupling strength $U$ relies on controlling the wave-function 
overlap between the two internal states in a state-dependent microwave potential~\cite{Riedel+:10}. The 
main advantage of the latter approach is that it also works in magnetic traps and in the absence of a convenient Feshbach resonance for the relevant pair of internal atomic states.

The Hamiltonian of the system under consideration [cf. Eq.~\eqref{eq:HamiltonianIBJJ}]
is given by the sum of the nonlinear OAT term $\hbar\chi J_z^2$ and the Rabi-coupling (turning) term 
$-\hbar\Omega(t) J_x$; the latter describes a time-dependent rotation around the $J_x$ axis, with the rotation rate $\Omega(t)$. Therefore, this Hamiltonian describes modified -- due to the presence of time-dependent $\Omega(t)$ -- TNT-type dynamics of spin squeezing in internal BJJs. The dimensionless parameter $\Lambda \equiv N\chi/\Omega$ is conventionally 
used to quantify the relative importance of the nonlinear- and Rabi couplings in an internal BJJ~\cite{Gross:12}. Because in the problem at hand we assumed that Rabi coupling is time-dependent [i.e. $\Omega=\Omega(t)$], the parameter $\Lambda$ will also have a nontrivial time-dependence in what follows. 

Generally speaking, we can distinguish three different regimes 
for a system described by the Hamiltonian in Eq.~\eqref{eq:HamiltonianIBJJ}. 
The Rabi regime corresponds to the noninteracting limit ($\Lambda \ll 1$) of such a system, where its ground state is a CSS with maximal mean collective-spin length $\langle J_{x}\rangle = N/2$ and equal
spin fluctuations in the orthogonal directions (i.e. variances $\Delta J_{y}^2 = \Delta J_{z}^2 = N/4$). The latter is a special case of more 
general CSSs $|\theta,\phi\rangle\equiv e^{-i\phi J_z} e^{-i\theta J_y}|J,J\rangle$, where $\theta$ and $\phi$ are the spherical polar angles ($0\leq \theta< \pi$, $ 0\leq \phi< 2\pi$) and $|J,J\rangle$ is the joint eigenstate of $J^2$ and $J_z$ with the highest possible value (equal to $J=N/2$) of $J_z$. This state corresponds to 
all $N$ atoms being in the same single-particle state -- namely, 
an equal linear combination of the two modes $|\psi_1\rangle$ and $|\psi_2\rangle$ -- and is characterized by the complete absence of quantum correlations between particles; it is given by
\begin{equation} \label{CSSdef}
|\theta=\pi/2,\phi=0\rangle=\frac{1}{\sqrt{2^N N!}}\:( a^{\dagger}_1 
+ a^{\dagger}_2)^{N}|\textrm{vac}\rangle \:,
\end{equation}
where $|\textrm{vac}\rangle$ is the vacuum state. 

In the presence of increased interactions ($1<\Lambda< N$), fluctuations in the relative atom number in the 
two modes -- which translates to fluctuations in $ J_{z}$ -- become energetically unfavorable. This is 
the Josephson regime, where the ground state is a coherent spin-squeezed state, in which the spin fluctuations 
in the $ J_{z}$ direction are reduced at the expense of increased fluctuations in the $ J_{y}$ direction
and reduced mean collective-spin length. Finally, in the strongly-interacting case (Fock regime), i.e. for $\Lambda 
\gg N$, the ground state of the system is a strongly spin-squeezed state with vanishing mean collective-spin length. 

It is pertinent to note that the Hamiltonian 
of the system  under consideration [cf. Eq.~\eqref{eq:HamiltonianIBJJ}] is invariant under the exchange of the single-boson modes $|\psi_1\rangle$ and $|\psi_2\rangle$. Namely, under the transformation $a_1\leftrightarrows a_2$,
we have that $J_x\rightarrow J_x$ and 
$J_z\rightarrow -J_z$, thus the Hamiltonian 
in Eq.~\eqref{eq:HamiltonianIBJJ} remains unchanged. Therefore, the system under consideration possesses a parity symmetry,
which in turn guarantees the symmetry-protected adiabatic evolution~\cite{Zhuang+:20}. Indeed, in the $\chi>0$ case discussed here one possible route towards generating spin-squeezed states is the adiabatic evolution. 

In the following, we set out to determine the time-dependence $\Omega(t)$ of the linear-coupling strength [or, equivalently, 
that of the parameter $\Lambda(t)$], that leads to the evolution of the system at hand from an initial state to a final state 
with a larger corresponding spin squeezing. More precisely, our choice of the initial parameters of the system 
will correspond to the Josephson regime (i.e. $1 < \Lambda_0 < N$), thus the relevant ground state -- the initial state 
of our control scheme -- will already be a spin-squeezed state. However, the objective of our scheme is to steer the 
system towards a final state characterized by a much larger spin squeezing. In what follows, this goal will be achieved 
using the STA and eSTA approaches (see Sec. III below).

\subsection{Figures of merit for spin squeezing} \label{SpinSqueezingParams}
Because in the present work we will be concerned with the preparation of spin-squeezed states, it is pertinent to introduce 
the figures of merit that can be used to quantify spin squeezing in the system at hand. Two important figures of merit
that can be used for this purpose are the number-squeezing- and coherent spin-squeezing parameters.

The number-squeezing parameter, also known as the Kitagawa-Ueda spin-squeezing parameter, is defined as~\cite{Wineland+:92}
\begin{equation}\label{NumberSqueezing}
\xi^2_{N}(t) = \frac{\Delta  J^2_z}{(\Delta J_z^2)_{\text{bin}}}=
\frac{\Delta  J^2_z}{N/4} \:,
\end{equation}
with $\Delta  J^2_z \equiv \langle  J^2_z \rangle - \langle  J_z \rangle^2$ being the variance of
the operator $J_z$, i.e. we are considering the squeezing in spin $z$-direction, where a smaller value of $\xi^2_{N}(t)$ corresponds to larger squeezing. The time argument on the left-hand-side of Eq.~\eqref{NumberSqueezing} reflects the fact that in the time-dependent state-preparation problem under consideration the parameter $\xi^2_{N}$ also depends 
on time. In Eq.~\eqref{NumberSqueezing}, $(\Delta J_z^2)_{\text{bin}}=J/2 = N/4$ (the shot-noise limit) corresponds to
the reference CSS.

The coherent spin-squeezing parameter, which is often referred to as the Wineland parameter and in the problem at hand is time dependent, is given by~\cite{Sorensen+:01}
\begin{equation}\label{CoherentSpinSqueezing}
\xi^2_{S}(t) = \frac{N\Delta  J^2_z}{\langle
 J_x \rangle^2} = \frac{\xi^2_{N}(t)}{\zeta^2(t)} \:,
\end{equation}
where $\zeta (t)\equiv\langle \Psi(t)|2 J_x/N|\Psi(t)\rangle$ quantifies the phase coherence of the many-body state $|\Psi(t)\rangle$. The parameter $\xi_{S}$ serves to characterize the interplay between an improvement in number squeezing and loss of coherence; it can be used to quantify precision gain in interferometry, given that for spin-squeezed states the interferometric precision is increased to $\Delta\Theta=\xi_{S}/\sqrt{N}$~\cite{Gross:12}. Furthermore, whenever a many-body state satisfies the inequality $\xi^2_{S}<1$ the state in question is entangled~\cite{Sorensen+:01}.

\section{Improved STA scheme and application of eSTA formalism} \label{ApplicationESTA}
In this section, we will present three different approaches to obtain the control function $\Omega(t)$, 
i.e. the time-dependence of the linear-coupling strength, with the goal of steering the system under 
consideration from the initial ground state at $t=0$ for a given $\Omega_0 \equiv \Omega(t=0)$ to the 
previously specified, strongly spin-squeezed, target state at $t=t_f$, i.e. the ground state of the 
total Hamiltonian of the system  for $\Omega_f\equiv \Omega(t=t_f)$.

Before providing the essential details of the three aforementioned control schemes for engineering spin-squeezed states in the system at hand, we first briefly review their common ingredient - the well-known mapping of the system described by Eq. \ref{eq:HamiltonianIBJJ} to the continuum (see Sec.~\ref{MapContinuum} below). We then brefly review the well-established STA-based control scheme of Ref.~\cite{yuste2013}; the corresponding time-dependence of the linear-coupling strength will be denoted by $\Omega_1(t)$ [Sec.~\ref{STAschemeDescr}]. We follow this up with the presentation of an improved STA control scheme; the latter results 
in the control function $\Omega_2(t)$[Sec.~\ref{impSTAschemeDescr}].
Finally, we describe how the eSTA formalism, proposed and developed by Whitty {\em et al.} in Ref.~\cite{Whitty+:20}, can be employed in the quantum-state engineering problem at hand; this last approach leads to the time dependence $\Omega_e(t)$ of the linear-coupling strength [Sec.~\ref{eSTAschemeDescr}].

\subsection{Mapping to the continuum} \label{MapContinuum}
It is well-known that the two-site (or two-state)
Bose-Hubbard model of a BJJ can be mapped to a Schr\"odinger equation in the continuum with approximately a harmonic potential \cite{Mahmud+:05,Shchesnovich+Trippenbach:08}.
We will briefly review the main details of this mapping, which will be used frequently in the remainder of this paper.

With $ \{\ket{m}\} $ ($ m = -N/2, ..., N/2 $) being the eigenvalues of the operator $ J_{z}$, the general state $ \ket{\Psi}$ of a BJJ can be written as
\begin{equation}\label{eq:GeneralStateAngularMomentum}
\ket{\Psi} = \sum_{m = -N/2}^{N/2} c_{m}\ket{m} \:.
\end{equation}
The coefficients $c_{m}$ in this expansion ought to satisfy the coupled equations
\begin{eqnarray}
	\lefteqn{i \hbar \frac{d}{dt}c_{m}(t)} && \nonumber\\
	&=&  - \frac{\hbar \Omega (t)}{2} \left[ \beta_{m} c_{m+1}(t) + \beta_{m-1}c_{m-1}(t) \right]
	+ \hbar\chi m^2 c_{m}(t) \:,\nonumber\\
	\label{eq:numerical}
\end{eqnarray}
where
$\beta_m = \sqrt{\left(\frac{N}{2} + m +1\right)	\left(\frac{N}{2} -m\right)}$.
We define now $h=\frac{1}{N/2}$ and $z_m = \frac{m}{N/2}$ (the relative population difference between the two relevant hyperfine states). We also switch to the continuum version for the relative 
population difference ($z_m\rightarrow z$), introducing at the same time a dimensionless time 
$\tau$, such that $t= \tau / \chi$. In this manner, we can recast Eq.~\eqref{eq:numerical} 
as
\begin{eqnarray}
\lefteqn{i h \frac{\partial}{\partial \tau} \psi(\tau,z)} & & \nonumber\\
	&=& - \frac{\Omega(t)}{\chi} \left[ b_{h}(z-h)\psi(\tau, z-h) + b_{h}(z)\psi(\tau, z+h)\right] \nonumber\\
	&  & +  \frac{N}{2} z^2 \psi(\tau, z)\nonumber\\
 &=& H_S \psi(\tau,z)
 \label{eq:hamexact}
\end{eqnarray}
with
\begin{eqnarray}
b_{h} (z) = \frac{1}{2}\sqrt{\left(1 + z +h\right)
\left(1 -z \right)} \:, \label{eq:CoefficientCollectingN}
\end{eqnarray}
and
\begin{eqnarray}
H_S = - \frac{\Omega(t)}{\chi} \left[ e^{-i h \partial_z} b_{h}(z) + b_{h}(z)e^{i h \partial z} \right]
+ \frac{N}{2} z^2. \label{ham_S}
\end{eqnarray}
We set at this point $b_h (z) = 0$ for all $z \le -1-h$ and $z \ge 1$, as well as $\psi (t,z) = 0$ for $z \le -1-h$ and $z \ge 1+h$. 
It is worthwhile noting that if this equation is satisfied for $z \in [-1-h,1+h]$ (note that it is trivially satisfied outside of 
this interval) then it is evidently also satisfied for all $z_m$ ($m=-N/2,...N/2$) in the previously employed discrete description. 

It is interesting to note that if one rewrites the above equation with a different dimensionless time $\bar\tau=N \tau$, then the above equation using $\bar\tau$ would only depend on the dimensionless quantity $1/\Lambda(t)= \frac{\Omega(t)}{N \chi}$. We will come back to this fact at a later point in the discussion.

We now proceed to derive an approximated version of Eq.~\eqref{eq:hamexact}.
For small $h$ and by assuming that $z$ is small (i.e. the population difference between  the two states is small compared to the total particle number $N$) and neglecting a 
constant energy shift, we obtain a Schr\"odinger equation with an approximated
Hamiltonian of a harmonic oscillator
\begin{equation}\label{ham_h1}
H_{1} = -h^2 \frac{\Omega(t)}{2\chi}\frac{\partial^2}{\partial z^2} + \left[\frac{N}{2} + \frac{\Omega(t)}{2\chi}\right] z^2 \:.
\end{equation}
We can make an additional approximation by assuming that $\frac{N}{2} \gg \frac{\Omega(t)}{2\chi}$, which leads to
\begin{equation}\label{ham_ho}
H_0 = -h^2 \frac{\Omega(t)}{2\chi}\frac{\partial^2}{\partial z^2} + \frac{N}{2} z^2 \:.
\end{equation}
Note that this last Hamiltonian has been the point of departure 
for deriving the STA scheme in Ref.~\cite{yuste2013}.

\subsection{STA scheme for harmonic approximation $H_{0}$} \label{STAschemeDescr}
The STA formalism for the case of internal BJJs is outlined in Ref.~\cite{yuste2013}. By employing Lewis-Riesenfeld invariants, we
can design a solution of the time-dependent
Schr\"odinger equation for the harmonic oscillator Hamiltonian $H_{0}$.
By making use of Fourier transform and Lewis-Riesenfeld 
invariants (for details, see Ref.~\cite{yuste2013} and Appendix~\ref{AppendMomSpace}), we find that the wave function 
\begin{eqnarray}
\chi_n(\tau,z) &=& 
\frac{\sqrt{b(\tau)}}{\pi^{1/4} \sqrt{2^n n!}}
\frac{i^n \sqrt{k_0}}{\sqrt{1-i \beta}}
\left(\frac{1+i\beta}{1-i\beta}\right)^{n/2} \nonumber\\
& & \times 
\fexp{i \varphi(\tau)
- \frac{z^2 b(\tau)^2}{2}
\frac{k_0^2}{1-i\beta}}
\nonumber\\
& & \times	H_{n}\left[\frac{z b(\tau) k_0}{\sqrt{1+\beta^2}}\right] \:, % Hermite polynomial term
	\label{eq:chi}
\end{eqnarray}
satisfies the time-dependent Schr\"odinger equation for 
the harmonic-oscillator Hamiltonian $H_{0}$ [cf. Eq.~\eqref{ham_ho}]. Here $k_0 = \frac{N}{2\sqrt{C}}$ 
and $\beta=\frac{1}{2C} b \dot b$, where $C=\sqrt{\frac{\Omega_0 N}{4\chi}} = \frac{N}{2\sqrt{\Lambda_0}} $ and the auxiliary function $b(\tau)$ must be a solution of the Ermakov equation
\begin{eqnarray}
b''(\tau) - \frac{N^2}{\Lambda_0 b(\tau)^3} + b(\tau) \frac{\Omega(\tau)}{\Omega_0} \frac{N^2}{\Lambda_0} = 0 \:.
\end{eqnarray}
$\varphi(t)$ in Eq.~\eqref{eq:chi} stands for 
$\varphi(t) = -\int_0^t ds\, (1+2n) \frac{C}{b(s)^2}$.
We now employ inverse engineering to first fix a function $b(\tau)$ that satisfies the boundary conditions 
$b(0)=1$, $b(\tau_f)=\sqrt[4]{\Omega_0/\Omega_f}$, $b'(0)=b'(\tau_f)=b''(0)=b''(\tau_f)=0$.
We choose here a polynomial of degree $6$ that satisfies these conditions. By inverting the above equation, we obtain an explicit expression for the sought-after physical control function 
$\Omega(\tau)$:
\begin{eqnarray}
\Omega(\tau) = \Omega_0
\left[\frac{1}{b(\tau)^4} - \frac{\Lambda_0}{N^2} \frac{1}{b(\tau)} \frac{\partial^2 b}{\partial\tau^2} \right]\:,
\label{eq:Omega}
\end{eqnarray}
where $\Lambda_0 \equiv \chi N/\Omega_0$.
We will call this the first STA scheme in the remainder of this paper and 
we denote the corresponding control function with $\Omega_1$.

\subsection{Improved STA scheme for harmonic approximation $H_{0}$} \label{impSTAschemeDescr}
We now proceed to introduce an improved STA scheme. The idea is that while the above approximation is still used for the time evolution, we consider the exact initial and final states without this approximation for designing the boundary conditions for the auxiliary function $b(\tau)$ [cf. Sec.~\ref{STAschemeDescr}]. To be more specific, we calculate the exact ground state of the Hamiltonian \eqref{ham_h1} at initial and final time.
For the ground state at initial time $\tau=0$, we obtain a function of the form of \eqref{eq:chi} with
\begin{eqnarray}
b(\tau) = \sqrt[4]{1 + 1/\Lambda_0}\:, \quad 
b'(0) = 0 \:.
\label{eq:imp0}
\end{eqnarray}
For the ground state at final time $\tau=\tau_f$, we get the function of the form of \eqref{eq:chi} with
\begin{eqnarray}
   b(\tau_f) = \left[\frac{\Omega_0}{\Omega_f}
   \left(1 + \frac{\Omega_f}{\Omega_0 \Lambda_0}
   \right)\right]^{1/4}\:,\quad b'(\tau_f) = 0 \:.
   \label{eq:impf}
\end{eqnarray}
We choose now a polynomial satisfying the boundary conditions \eqref{eq:imp0} and \eqref{eq:impf}. In addition, we demand that \eqref{eq:Omega} is also satisfied at the initial- and final times with these new boundary conditions. This leads to two additional boundary conditions
\begin{eqnarray}
b''(0) &=& - \frac{N^2}{\Lambda_0^2} \frac{1}{\left(1 + 1/\Lambda_0\right)^{3/4}}\:,\nonumber\\
b''(\tau_f) &=& - \frac{N^2}{\Lambda_0^2} \frac{\Omega_f/\Omega_0}{\left(\Omega_0/\Omega_f + 1/\Lambda_0
\right)^{3/4}} \:.
\label{eq:imp2}
 \end{eqnarray}
We choose here again a polynomial of degree $6$ that satisfies the fix conditions \eqref{eq:imp0},\eqref{eq:impf} and \eqref{eq:imp2}.
We will then use \eqref{eq:Omega} with this polynomial to calculate the control function. We will call this the improved STA scheme in the following and will denote the corresponding control function with $\Omega_2$.

\subsection{eSTA-corrected control function}\label{eSTAschemeDescr}
We will now derive an eSTA correction to the improved STA scheme from the last subsection. The principal idea behind 
the eSTA formalism is to modify the STA control function by taking into account that the approximated Hamiltonian 
$H_0$ is different from the exact system Hamiltonian $H_S$ [cf. Eq.~\eqref{ham_ho}].
Let $\Delta H = H_S - H_0$ be the difference between these two Hamiltonians and $\vec\lambda = (\lambda_1, \lambda_2)$. We consider now the modified control function $\Omega_e (\tau) = \Omega_2 (\tau) + P_{\vec\lambda}(\tau)$, where $P_{\vec{\lambda}}(\tau)$ is a polynomial of degree $4$ that fulfills the following conditions:
\begin{eqnarray}
P_{\vec{\lambda}}(0)=P_{\vec{\lambda}}(\tau_f)=0, \nonumber\\
P_{\vec{\lambda}}(\tau_f/3) = \lambda_1, P_{\vec{\lambda}}(2\tau_f/3) = \lambda_2 \:.
\end{eqnarray}
These corrections $\vec{\lambda}$ are calculated by making use of the eSTA formalism~\cite{Whitty+:20, Whitty2022b}.

To begin with, we define the auxiliary functions $G_n$, $K_{n}$, and $\mathrm{H}$ that will be used to evaluate $\vec{\lambda}$.
In terms of the STA invariant eigenfunctions $\chi_{n}$ 
of the approximated Hamiltonian [cf. Eq. \eqref{eq:chi}] the scalar auxiliary function $ G_n $ is given by~\cite{Whitty+:20}
\begin{equation}
G_{n} = \int_{0}^{t_{f}} dt \braket{\chi_n}{\Delta H |\chi_{0}} \:. \label{eq:Gn}
\end{equation}
The correspionding expression for the vector auxiliary function $\vec{K}_{n} $ 
reads~\cite{Whitty+:20} 
\begin{eqnarray}
\vec{K}_{n} = \int_{0}^{t_{f}} dt \braket{\chi_{n}}{\vec{\nabla} H_S|\chi_{0}} \:, \label{eq:Kn}
\end{eqnarray}
with $\vec{\nabla} H_S$ being the gradient of the Hamiltonian $H_S$ with respect to the control 
parameter $\vec\lambda$. Finally, the matrix elements of $\mathrm{H}$ are given by~\cite{Whitty+:20}
\begin{eqnarray}
\mathrm{H}_{l,k} = \sum_{n = 1}^{\mathcal{N}}\left[G_{n}(W_{n})_{l,k} + \left(  \vec{K}^{*}_{n}
\right)_{k}\left(  \vec{K}_{n} \right)_{l}\right] \:,\label{eq:HessianMatrix}
\end{eqnarray}
where $ (W_{n})_{l,k} = \braket{\chi_n}{\partial_{\lambda_l}\partial_{\lambda_k} H_S | \chi_0}$ stands for a matrix of second derivatives with respect to the control parameter; here $\mathcal{N}$ is the number of STA invariant eigenfunctions that are taken into account.

Having defined the relevant auxiliary functions, we could compute the correction parameters $\vec\lambda$ 
using $G_n$ and $\vec{K}_{n}$, assuming that a unit fidelity can be achieved for the exact 
Hamiltonian~\cite{Whitty+:20}.
The same correction parameters can be obtained in an alternate fashion -- namely, by employing the above expressions 
for $G_n$, $\mathbf{K}_n$, and $\mathrm{H}_{l,k}$~\cite{Whitty2022b}. By making use of the latter  procedure, the correction 
parameters $\vec\lambda$ are given by
\begin{eqnarray}
\vec\lambda = -\frac{\norm{\vec{v}}^{2}}
{\vec{v}^{T}\mathrm{H}\vec{v}}\:\vec{v} \:,
	\label{eq:eSTAcorrectionsHessian}
	\label{esta_labda}
\end{eqnarray}
where
\begin{eqnarray}
	\vec{v} = \sum_{n=1}^{\mathcal{N}}\text{Re}(G_{n}^{*}\vec{K}_{n})\:.
	\label{eq:FidelityVector}
\end{eqnarray}

Here, we have now $H_S$ given by Eq. \eqref{ham_S} and $H_0$ given by Eq. (\ref{ham_ho}). Therefore, the resulting expression for $\Delta H$ reads as
\begin{equation}
	\Delta H = - \frac{\Omega(t)}{\chi} \left[ e^{-i h \partial_z} b_{h}(z) + b_{h}(z)e^{i h \partial z} 
 + \frac{h^2}{2}\frac{\partial^2}{\partial z^2} \right] \:.
 \label{eq:D}
\end{equation}

Note that $\Omega_s (\tau)$ with the above defined $P_{\vec{\lambda}}$ is only linear in $\vec\lambda$, and therefore $H_s$ is only linear in $\vec\lambda$ such that it follows $W_{n} = 0$.
In this approach, the correction parameters 
$\vec\lambda$ are thus given by Eq.~\eqref{esta_labda}. 
In what follows, we set $\mathcal{N} = 2$ and denote the resulting 
control function by $\Omega_e (t)$.

% --------------------------------------------------------------------------------------------------
% Results and Discussion
% -----------------------------------------------------------------------------------------------

\section{Results and discussion\label{sect_results}}
In the following, we discuss and compare the results for the spin-squeezing parameters and the target-state fidelity obtained using the original STA-based scheme, the improved STA scheme, and the newly proposed eSTA-based scheme [cf. Sec.~\ref{ApplicationESTA}];
the corresponding time-dependent linear-coupling strengths (control functions) in these three control schemes are denoted by $\Omega_1(t)$, $\Omega_2(t)$, and $\Omega_e (t)$, respectively. For the sake of completeness, we also compare the obtained results with those resulting from a simple linear adiabatic sweep, characterized by the time dependence $\Omega_{\textrm{ad}} (t) = \Omega_0 + (\Omega_f - \Omega_0)t/t_f $ [cf. Sec.~\ref{IntroIBJJ}] of the control function.

For better comparison with existing experiments, in the following we will also switch back from the dimensionless variables used in this section to the general dimensional quantities. In addition, we will use $1/\chi$ as the characteristic time scale; note that $\Lambda(t) = N\chi/\Omega(t)$. In what follows, we will always use the ratio $\Omega_f/\Omega_0 = 1/10$ resp. $\Lambda_f/\Lambda_0 = 10$.

We start the discussion of obtained results by presenting results for the target-state fidelity and spin squeezing parameters 
(Sec.~\ref{sect_fidelity}), followed by 
a short discussion of the robustness of
STA- and eSTA-based schemes (Sec.~\ref{StabilityControlScheme}), and, finally, compare our results with the existing experiments and discuss future prospects (Sec.~\ref{ExpCompareProspects}).

\subsection{Target-state fidelity and spin squeezing}\label{sect_fidelity}
We start by discussing the results for the target-state fidelity obtained using different control schemes. This fidelity is defined as $F = \left| \langle \Psi_{\textrm{T}} | \Psi(t_f) \rangle \right|^2$,
where $\Psi_{\textrm{T}}$ is the target spin-squeezed state [i.e. the ground state of the Hamiltonian $H_{\textrm{IBJJ}}$ in Eq.~\eqref{eq:HamiltonianIBJJ}
for $t = t_f$] and $\ket{\Psi(t_f)}$ the actual 
state of the system at $t = t_f$ obtained through the time evolution governed by the Hamiltonian $H_{\textrm{IBJJ}}(t)$ [cf. Eq.~\eqref{eq:HamiltonianIBJJ}], whose time dependence originates from the time-dependent linear-coupling strength $\Omega(t)$. We will consider different particle numbers $N$ and initial values $\Omega_0$ resp. $\Lambda_0$ in the following.

%
%
% ------------------------ Fig. 2 -------------------------------
\begin{figure}
	\begin{center}
	\includegraphics[width=\linewidth]{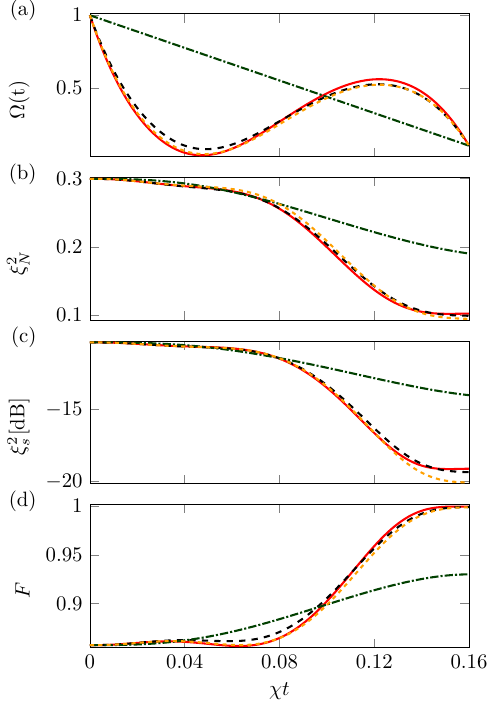}
	\end{center}
	\caption{Time evolution for $N=50$ of (a) linear coupling strength (control function) $\Omega (t)$,
(b) number-squeezing parameter $\xi_N^2(t)$,
		(c) coherent-spin-squeezing parameter $\xi_S^2 (t)$ in dB, (d) 
        target-state fidelity $F(t)$.
The results shown here correspond to eSTA ($\Omega_e$: red, solid lines), improved STA ($\Omega_2$, orange, small dashed lines), first STA scheme ($\Omega_1$, black, dashed lines) 
and as a reference linear adiabatic scheme ($\Omega_{\textrm{ad}}$, green, dashed-dotted lines).
The parameter values used are $\chi t_f = 0.16$, $\Lambda_0 = 10$, $\Omega_f/\Omega_0 = 0.1$.
\label{fig:example50}}
\end{figure}
% -------------------------------------------------

%
%
% ------------------------ Fig. 3 -------------------------------
\begin{figure}
	\begin{center}
	\includegraphics[width=\linewidth]{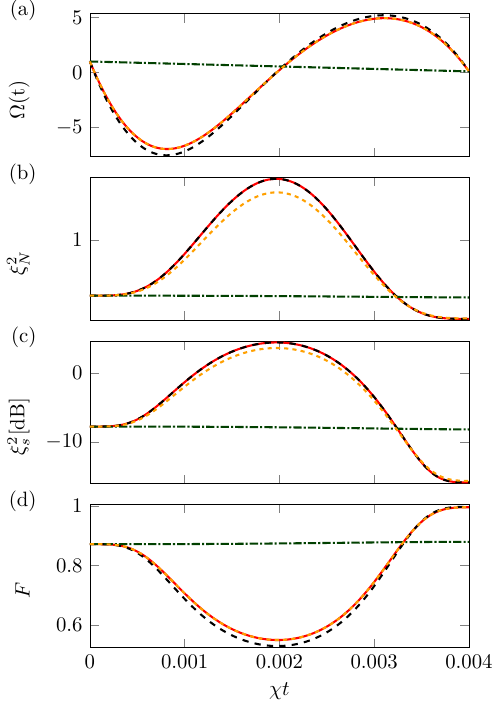}
	\end{center}
	\caption{Time evolution for $N=400$ of (a) linear coupling strength (control function) $\Omega (t)$,
(b) number-squeezing parameter        $\xi_N^2(t)$,
		(c) coherent-spin-squeezing parameter $\xi_S^2 (t)$ in dB, (d) target-state fidelity $F(t)$. The results shown here correspond to eSTA ($\Omega_e$:
red, solid lines), improved STA ($\Omega_2$, orange, small dashed lines), first STA scheme ($\Omega_1$, black, dashed lines) 
and as a reference linear adiabatic scheme ($\Omega_{\textrm{ad}}$, green, dashed-dotted lines).
The parameter values used are $\chi t_f = 0.004$, $\Lambda_0 = 10$, $\Omega_f/\Omega_0 = 0.1$.
\label{fig:example400}}
\end{figure}
% -------------------------------------------------

As an example, the time evolution of different quantifies with $N=50$ particles and the final time $\chi t_f = 0.16$ is shown in Fig.~\ref{fig:example50} as well as with $N=400$ particles and the final time $\chi t_f = 0.004$ is shown in Fig.~\ref{fig:example400}. In detail, in Fig.~\ref{fig:example50}(a) resp. Fig.~\ref{fig:example400}(a) we can see the different control functions. We compare here the adiabatic control $\Omega_{\textrm{ad}} (t)$, original, the STA scheme $\Omega_1 (t)$,
the improved STA scheme $\Omega_2 (t)$
and the eSTA scheme $\Omega_e (t)$.
Note that, in principle, formula \eqref{eq:Omega} could result in a negative value of the control function $\Omega$ for some time; however, this is not a problem as the existing experimental capabilities allow also rapid changes in the phase of $\Omega$ and therefore also for negative values of $\Omega(t)$ \cite{Muessel+:15}.

The remaining figures 
illustrate the time evolution of the number squeezing $\xi_N^2$ \eqref{NumberSqueezing} [Fig.~\ref{fig:example50}(b) resp. \ref{fig:example400}(b)] and the 
coherent spin-squeezing parameter $\xi_{S}^2$ \eqref{CoherentSpinSqueezing} expressed in dB [Fig.~\ref{fig:example50}(c) resp. Fig.~\ref{fig:example50}(c)] 
as well as the fidelity $F$ [Fig. \ref{fig:example50}(d) resp. Fig. \ref{fig:example400}(d)] when the four different 
control functions are applied.

% ------------------------ Fig. 4 -------------------------------
\begin{figure*}[t]
	\begin{center}
	\includegraphics[width=0.95\linewidth]{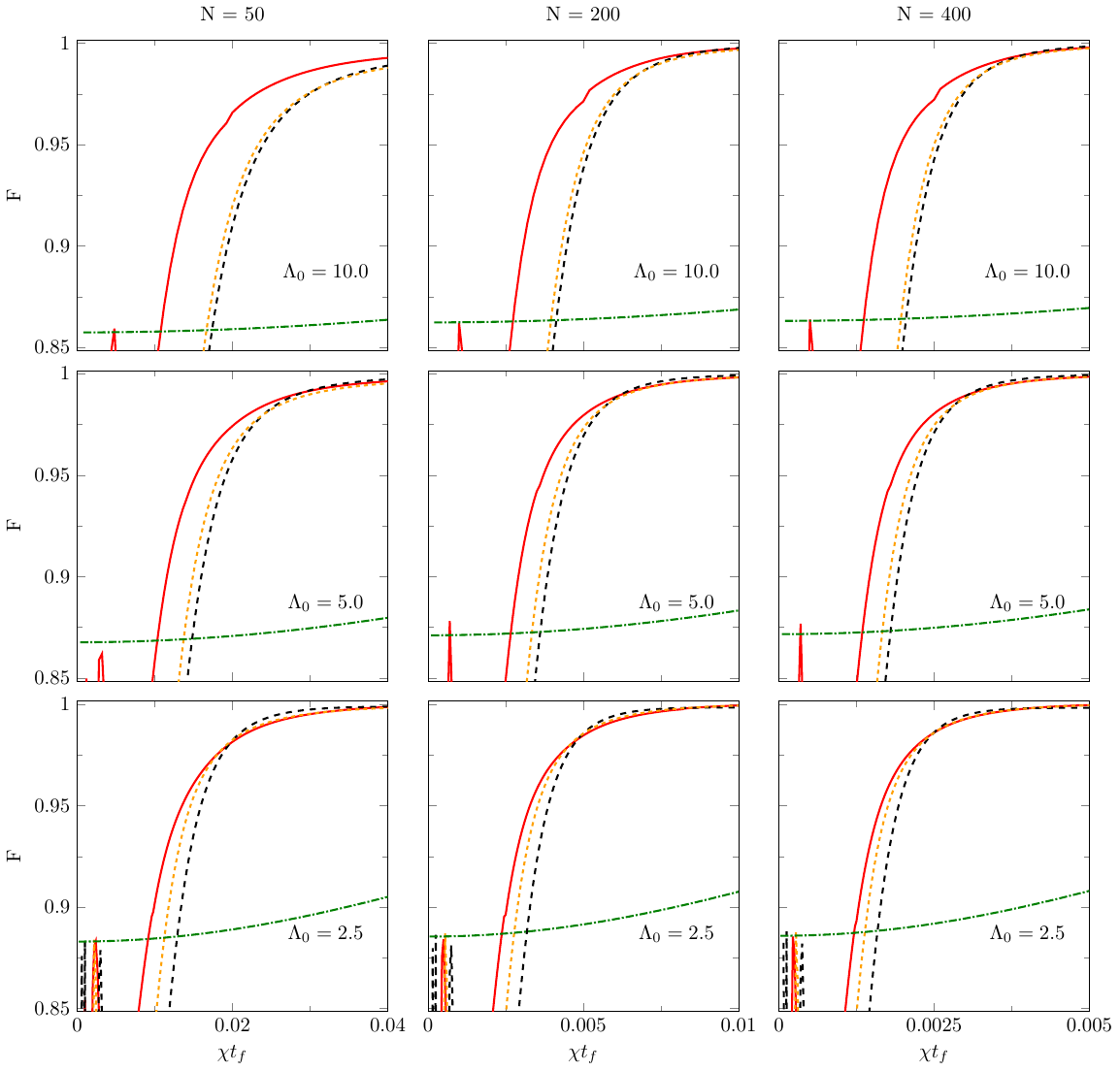}
	\end{center}
	\caption{Target-state fidelities $F$ versus total evolution time $t_f$ for the different control schemes: eSTA ($\Omega_e$:
red, solid lines), STA ($\Omega_1$, black, dashed lines), improved STA ($\Omega_2$, orange, small dashed lines), and, as a reference, linear adiabatic scheme ($\Omega_{\textrm{ad}}$, green, dashed-dotted lines) for different particle numbers
$N=50$ (left column), $N=200$ (middle column), $N=400$ (right column) as well as different initial values of $\Lambda_0$ resp. $\Omega_0=\chi N/\Lambda_0$: $\Lambda_0 = 10.0$ (first row), $\Lambda_0 = 5.0$ (second row), $\Lambda_0 = 2.5$ (third row). The results shown here correspond to $\Omega_f/\Omega_0 = \Lambda_0/\Lambda_f = 1/10$.
		\label{fig:fidelity}}
\end{figure*}
% ---------------------------------------------------------------

The results obtained for the target-state fidelity,within four different control schemes (respectively 
characterized by the control functions $\Omega_{\textrm{ad}}$, $\Omega_1$, $\Omega_2$, $\Omega_e$) and 
for varying total times $t_f$, are illustrated in Fig.~\ref{fig:fidelity}; the results are displayed for 
$N = 50$, $200$, and $400$ particles. From these results we can infer that generally the eSTA-based 
protocol outperforms its STA-based counterpart, especially for shorter times $t_f$. Another important 
feature that is worthwhile pointing out is the fact that the improved version of the STA-based protocol 
(characterized by $\Omega_2$) is consistently better than the conventional STA ($\Omega_1$). This result 
shows improvements can be achieved even without applying the eSTA formalism, but using the improved
boundary conditions for the system under consideration. Moreover, the fidelity of the three protocols 
tend to converge to 1 as the final time increases. It is also interesting to see how increasing the 
interaction strength $\chi$ (i.e. moving up a column in the figure grid), corresponds to an improvement 
in the performance of the protocols. This can be explained by the fact that the approximation $\Lambda_0 
\gg 1$ gets increasingly more accurate as the value of $\Lambda_0$ grows. On the other hand, we can see 
how keeping constant the interaction strength and increasing the number of particles (this amounts to 
move along a row on the figure grid) does not change the overall behaviour of the fidelity, only the 
intrinsic time of the system scales with the number of particles.

A direct connection between the phase-space structure of a nonlinear classical system and the ground-state entanglement characteristics of the corresponding quantum system was established in the past. In particular, it was demonstrated that a precipitous growth of ground-state entanglement occurs in nonlinear quantum systems that have a classical counterpart undergoing -- for the same choice of values of the relevant parameters -- a supercritical pitchfork 
bifurcation~\cite{Hines+:05}. Interestingly, such phenomena have already been discussed in the context of spin squeezing
in internal BJJs, more precisely in a system exhibiting the conventional TNT-type dynamics (with time-independent linear- and nonlinear coupling strengths)~\cite{Zibold+:10}.

In connection with a possible occurrence of bifurcations of the aforementioned type in our current setting -- where an STA- or eSTA-based control scheme is employed for the generation of spin-squeezed states (with a time-dependent linear-coupling
strength) -- a comment is in order here. It is worthwhile pointing out that in our STA- and eSTA-based schemes the system is in 
a ground state of the total Hamiltonian only at the very beginning 
($t=0$) and the end ($t=t_f$) of its evolution; however, at
intermediate times ($0<t<t_f$) -- at which the critical value of the dimensionless parameter 
$\Lambda$ (for which the aforementioned bifurcation takes place in the case of the conventional TNT-type dynamics) may be reached -- the system is not even in the instantaneous ground state of its total (time-dependent) Hamiltonian. Therefore, the standard scenario for the occurrence of bifurcations of this type cannot 
be realized in this case; yet, it is worthwhile investigating 
in the future whether a more general dynamical analogue of the latter can result from the dynamics engendered by our control scheme.

\subsection{Robustness of the STA- and eSTA-based control schemes} \label{StabilityControlScheme}
\begin{figure}[b]
	\begin{center}
	 \includegraphics[width = 0.9\linewidth]{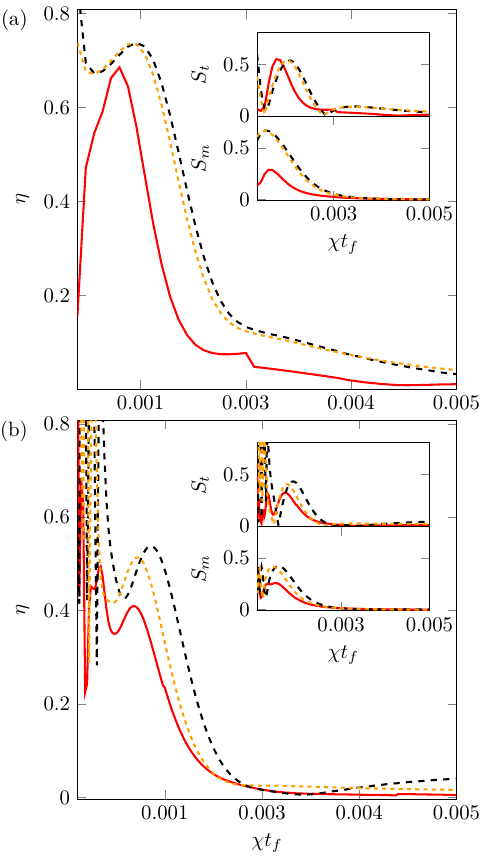}
	\end{center}
	\caption{Imperfection $\eta$ (outer figures) and sensitivities for systematic control-amplitude error 
 ($S_m$; lower inset) and systematic time-shift ($S_t$; upper inset) in a system with $N=400$ particles for 
 (a) $\Lambda_0=2.5$, (b) $\Lambda_0 = 10$. The results shown here correspond to the eSTA scheme ($\Omega_e$:red, solid lines), 
 STA scheme ($\Omega_1$, black, dashed lines), and improved STA scheme ($\Omega_2$, orange,small dashed lines).
 \label{fig:robustness}}
\end{figure}

It is important to ensure that a control scheme does not only provide a high fidelity but that it is also robust against errors. Therefore, we will now compare the robustness of the above control schemes against systematic errors, i.e. an unknown, constant error in the experimental setup. 
First, we consider a systematic error in the amplitude of the control function $\Omega (t)$ of the form $\Omega_{\delta,m}(t) = 
(1 + \delta) \Omega (t)$ for $ t \in [0, t_f]$ and for an unknown constant value of $\delta$.
To quantify the sensitivity of the control scheme to systematic errors of this type, we evaluate numerically the sensitivity
\begin{eqnarray}
S_m = \left|\frac{\partial F}{\partial \delta}\right|_{\delta=0} \:.
\label{eq_S}
\end{eqnarray}
for each control scheme used. Note that the lower this sensitivity is, the more stable the protocol is.  
The obtained results for $S_m$ are displayed in Fig.~\ref{fig:robustness} (lower inset) for $N=400$ and $\Lambda_0=5$ [see the 
lower inset of Fig.~\ref{fig:robustness}(a)] and $\Lambda_0=20$ [see the lower inset of Fig.~\ref{fig:robustness}(b)].

% ------------------------ Fig. 5 -------------------------------
% ---------------------------------------------------------------

We also consider a second type of systematic error. This second case is the systematic error in the time of the control function $\Omega (t)$ of the form
$\Omega_{\bar\delta,t} (t)= \Omega (t+t_f \bar\delta )$ for $t+t_f \bar\delta \in [0, t_f]$ and $\Omega_{\bar\delta,t} =
	\Omega (t)$ otherwise.
Similar to Eq.~\eqref{eq_S} above, we compute numerically the corresponding systematic error sensitivity
\begin{eqnarray}
S_t = \left|\frac{\partial F}{\partial \bar\delta}\right|_{\bar\delta=0} \:.
\end{eqnarray}
The results obtained for $S_t$ are illustrated in Fig.~\ref{fig:robustness} (upper inset) for $N=400$ and $\Lambda_0=5$ [see the
upper inset of Fig.~\ref{fig:robustness}(a)] and $\Lambda_0=20$ [see the upper inset of Fig.~\ref{fig:robustness}(b)].

It is pertinent to also incorporate the fidelity and these sensitivities in a single figure of merit,
which will be referred to as imperfection in what follows. This quantity is given by
\begin{equation}\label{eq:imperfection}
\eta = \sqrt{(1-F)^{2} + S_m^{2} + S_t^2} \:,
\end{equation}
where $F$ is the fidelity and $S_m, S_t$ are the sensitivities defined above. Therefore, a small value of $\eta$
corresponds to low infidelity (i.e. high fidelity) and small sensitivities (i.e. high degree of robustness) to 
the both systematic errors, i.e. the lower the value of $\eta$, the better the control scheme. The results are illustrated in Fig.~\ref{fig:robustness} (outer figures) for $N=400$ and $\Lambda_0=2.5$ [see Fig.~\ref{fig:robustness}(a)] and $\Lambda_0=10$ [see Fig.~\ref{fig:robustness}(b)]. What can be inferred from these results is that the best performance is achieved using the eSTA method.

\subsection{Comparison of various approaches for engineering spin-squeezed states in internal BJJs} \label{ExpCompareProspects}
In order to quantitatively assess our proposed eSTA-based scheme for engineering spin-squeezed states in internal BJJs, 
it is pertinent to compare the achievable spin squeezing obtained using this scheme with the ones obtained with previously known schemes~\cite{Gross+:10,Riedel+:10}. Generally speaking, for 
different total evolution times $t_f$, the eSTA-based scheme slightly outperforms its STA-based counterpart in terms of achievable spin squeezing, as quantified by the previously introduced parameters $\xi^2_{S}$ and $\xi^2_{N}$ [cf. Sec.~\ref{SpinSqueezingParams}]. However, the primary advantage of the eSTA approach, which makes it much more powerful than STA, lies in its superior robustness against systematic errors.

For the sake of illustration, it is instructive to consider a system with $N=400$ particles, with the same value of the nonlinear-coupling strength $\chi = 2\pi \times 0.063$\:Hz used in an experimental study of Ref.~\cite{Gross+:10}.
While in \cite{Gross+:10} the linear-coupling strength $\Omega$ was fixed to the constant value $\Omega=2\pi\times 2$\:Hz -- leading to a ratio $\chi/\Omega=0.03$ -- in the present work we consider a significant variation of the linear-coupling strength with time. As a result of this variation, the control parameter $\Lambda(t)$ is changed from $\Lambda(0)/N=\chi/\Omega(0)=1/160$ to $\Lambda(t_f)/N=\chi/\Omega(t_f)=1/16$ ($N=400$); in this manner, we demonstrate that our proposed eSTA-based scheme for the generation of spin-squeezed states is applicable even far away from the adiabatic regime.
Importantly, the total squeezed-state preparation time that we obtain here [cf. Fig.~\ref{fig:fidelity}, $N=400$] is then $t_f= 0.005\:\chi^{-1}\approx\:13$\:ms; this time is $35\%$ shorter than the squeezed-state preparation time of $20$\:ms found 
in Ref.~\cite{Gross+:10}.

Our numerical calculations indicate that the two schemes for engineering spin-squeezed states -- STA- and eSTA-based ones -- are quite comparable in terms of of achievable spin squeezing.
For example, in a system with $N = 50$ particles and $\Lambda_0/N = 0.2$ the eSTA control scheme allows one to attain a value of $ \xi_S^2 < - 18$\:dB while the adiabatic scheme only yields a value of $\xi_S^2 > - 14$\:dB, see Fig. \ref{fig:example50}.

The effects of the application of the eSTA protocol get increasingly less prominent as the number of particles increases. For instance, for a system with $N = 400$ particles, $t_f=0.1/\chi$ and $\Lambda_0/N = 10$ both the eSTA-based scheme and the STA-based one, yield $\xi_S^2 = -19$ dB. It is useful to recall that the eSTA protocol is designed with the aim to maximize the fidelity of the system. The fact that the squeezing obtained via eSTA protocol is comparable, if not better, than the one obtained using the STA approach is another argument in favor of eSTA. 

On the other hand, the eSTA-based protocol has a lower sensitivity $S_m$ than the STA and improved STA protocols against systematic errors as it can straightforwardly be inferred from the sensitivities shown in the insets of Fig. \ref{fig:robustness};
thus, the eSTA scheme is significantly less sensitive (i.e. more robust) to systematic errors in the amplitude of $\Omega$ than the STA-based one.
In the insets of the same figure, similar behavior is found for the systematic error pertaining to the timing of the control scheme $\Omega$; for the above choice of parameter values, the sensitivity $S_t $ is again lower for the eSTA scheme than for the STA scheme and its
improved version.

The observed superiority of the eSTA-based approach over its STA-based counterpart in the 
system under consideration is more prominent for shorter times $t_f$. Importantly, this superiority of the eSTA-based approach is not  dependent upon the particle number; it persists
even for a system with several hundred (or even thousand) particles. The fact that the eSTA method allows one to attain a rather strong spin squeezing in internal BJJs -- being at the same time much more robust against systematic errors than STA -- makes this method a highly promising candidate for the experimental realization of spin-squeezed states in this type of systems.

% ------------------------ Fig. 6 -------------------------------
\begin{figure}[t]
	\begin{center}
	 \includegraphics[width = 0.9\linewidth]{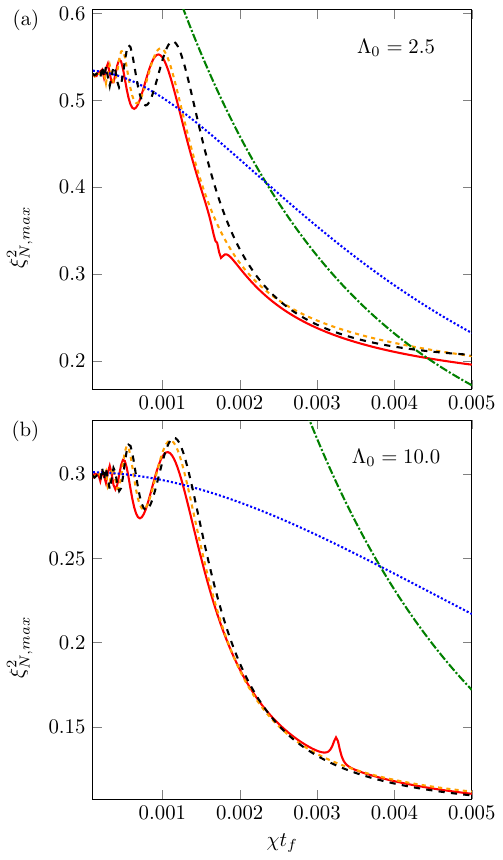}
	\end{center}
	\caption{Number-squeezing parameter $\xi_{N,max}^2(t_f)$ in the direction of maximal spin squeezing in the $y-z$ plane at final time $t_f$ in a system with $N=400$ particles, for (a) $\Lambda_0 = 2.5$, (b) $\Lambda_0 = 10.0$. The values of other parameters are the same as in Fig.~\ref{fig:fidelity}. The results shown here correspond to the following control schemes: eSTA ($\Omega_e$, red, solid lines), STA ($\Omega_1$, black, dashed lines), improved STA ($\Omega_2$, orange,small dashed lines), and  OAT, starting with the ground state pertaining to the value 
    $\Lambda_0$ of the parameter $\Lambda$ (blue line).
 For reference, the results obtained using the OAT scheme starting with CSS as the initial state are also displayed (green lines). \label{fig:oat}}
\end{figure}
% ---------------------------------------------------------------

It is pertinent to also compare our envisioned control schemes for the generation of spin-squeezed states with the well-known OAT approach~\cite{Kitagawa+Ueda:93}. More specifically yet, we consider here the number-squeezing parameter $\xi^2_{N,max}$ corresponding to the direction of maximal spin squeezing in the $y-z$ plane~\cite{Kitagawa+Ueda:93}. The obtained results for this particular figure of merit of spin squeezing are illustrated in Fig.~\ref{fig:oat}, where those obtained for the OAT scheme are compared with their counterparts pertaining to the STA- and eSTA schemes that 
correspond to the same initial state (blue line). What 
can be inferred from these results is that for short times the performance of the STA- and eSTA-based protocols is comparable to that of OAT, while for longer times (for $t_f \gtrsim 0.014\:\chi^{-1}$) the STA-based protocols achieve larger spin squeezing than the OAT-based one. It is important to note that the STA- and eSTA-based schemes are designed to achieve the maximal number squeezing in the corresponding target state, which represents the ground state of the total Hamiltonian of the system for given value $\Lambda_f$ 
of the dimensionless parameter $\Lambda$. For reference, the results obtained for the OAT scheme starting with the CSS state 
[cf. Eq.~\eqref{CSSdef}] as the initial state are also displayed in Fig.~\ref{fig:oat} (green line).

The TNT dynamics of spin squeezing in internal BJJs has so far been demonstrated experimentally by abruptly switching the nonlinear coupling to a finite value (i.e. by performing a nonlinear-coupling quench) in the presence of constant linear coupling $\Omega$~\cite{Muessel+:15}; in this case a Feshbach resonance was used as an enabling physical mechanism for increasing the nonlinear coupling strength. This investigated scenario of TNT dynamics, with a fixed ratio of the nonlinear- and linear coupling strenghts, has already also been shown -- in the absence of decoherence and losses -- to be locally optimal as far as the generation of spin squeezing is concerned~\cite{Sorelli+:19}. However, there are several reasons that make it quite plausible to expect that in the realistic experimental scenario (i.e. in the presence of decoherence and losses, as well as various experimental imperfections) the eSTA-based approach proposed here could yield comparable -- or, perhaps, even better -- results. To begin with, the eSTA method~\cite{Whitty+:20} has already been demonstrated to yield results approaching the relevant quantum speed limits in some other quantum-control problems~\cite{Hauck+Stojanovic:22,Whitty+:22}. In addition, the extraordinary robustness to various experimental imperfections that is inherent to the eSTA method -- compared to its parent STA methods and other control approaches -- also favors our envisioned approach. Finally, our scheme may turn out to be more robust to atomic losses (most prominently, two-body spin-relaxation losses for $F = 2$ hyperfine states); in the existing experimental realization~\cite{Muessel+:15}, for the same range of particle numbers as discussed here the combined effect of two- and three-body losses led to a substantial (above $20\%$) decrease of the maximal achievable spin squeezing.

\section{Conclusion and Outlook\label{SummConcl}}
In summary, in this paper we revisited -- from the quantum-control perspective -- the problem of robust, time-efficient engineering of spin-squeezed states in internal bosonic Josephson junctions with a time-dependent linear (Rabi) coupling.
Unlike earlier studies in which this state-engineering problem was treated using adiabatic- or STA methods, we addressed it here using the recently proposed eSTA approach. Taking the standard Lipkin-Meshkov-Glick-type Hamiltonian of this system -- which in addition to the standard one-axis twisting term includes the transverse-field (turning) term -- as our point of departure, we designed a robust eSTA-based state-preparation scheme. To characterize the twist-and-turn-type quantum dynamics underlying the preparation of spin-squeezed states in this system in a quantitative fashion, we computed the corresponding (time-dependent) target-state fidelity. We also evaluated two of the most relevant figures of merit of spin squeezing -- namely, the coherent spin-squeezing- and number-squeezing parameters -- in a broad range of the relevant parameters of the system.

Importantly, we showed that our eSTA-based scheme for engineering spin-squeezed states outperforms -- in terms of achievable state fidelities and the attendant state-preparation times -- not 
only the previously proposed STA protocols, but also their improved version; this superiority of our eSTA-based approach is not limited only to relatively small particle numbers, but persists even in systems containing several hundred particles. In particular, we demonstrated that the increased robustness of the eSTA approach, compared to its parent STA method, renders our proposed state-preparation scheme more amenable to experimental realizations than the previously suggested schemes for engineering spin-squeezed states in bosonic Josephson junctions. In order to further 
facilitate the envisioned experimental realizations, 
a future theoretical work should be devoted to discussing other possible decoherence effects (other than those already discussed here) in this system, most prominently that of atomic losses~\cite{Sinatra+:12}.

An interesting possible direction of future work pertains to harnessing the twist-and-turn type dynamics of the system we considered in the present work for generating metrologically useful non-Gaussian states~\cite{Strobel+:14,Sorelli+:19}. Such states are not amenable to the characterization by spin-squeezing parameters, requiring instead other entanglement witnesses, e.g. the quantum Fisher information~\cite{Pezze+:18}. Finally, another relevant direction for future study is to examine the domain of applicability of the two-mode model and describe the effects that cannot be captured by this model.

\begin{acknowledgments}
	V.M.S. acknowledges a useful discussion with P. Treutlein.
	M.O and A.R acknowledge that this publication has emanated from research
	supported in part by a research grant from Science Foundation Ireland (SFI)
	under Grant Number 19/FFP/6951.
	This research was also supported by the Deutsche Forschungsgemeinschaft
	(DFG) -- SFB 1119 -- 236615297 (V.M.S.).
\end{acknowledgments}

\appendix 

\section{Momentum-space eigenfunctions of Lewis-Riesenfeld invariant and their Fourier transforms} \label{AppendMomSpace}
In what follows, we provide mathematical details required to retrieve the Lewis-Riesenfeld invariant eigenfunctions in Eq.~\eqref{eq:chi}, starting from the harmonic-oscillator-like Hamiltonian of Eq.~\eqref{ham_ho}.

We begin by noting that in the Hamiltonian
of Eq.~\eqref{ham_ho}, the time-dependent control function $\Omega(t)$ appears in the prefactor to the kinetic term $\partial_z$, while there is a constant term $ N /2  $ associated with prefactor to the potential term $ z^{2} $. This Hamiltonian can be compared with the standard harmonic-oscillator Hamiltonian
\begin{equation} % Standard Harmonic Oscillator Hamiltonian
H_{\textrm{HO}} = \frac{1}{2}m\omega^{2}(t)x^{2} - \frac{\hbar^{2}}{2m} \partial_x^{2}~\:,
\label{eq:HamiltonianHarmonicOscillator}
\end{equation}
to which the formalism of Lewis-Riesenfeld invariants has been implemented. In the particlar case of the Hamiltonian in Eq.~\eqref{ham_ho} the control function $ \omega=\omega(t)$ is paired with the multiplicative position operator $ x $.
For the Hamiltonian \eqref{eq:HamiltonianHarmonicOscillator}, we can use the well-established invariant-based STA approach to obtain an invariant and its corresponding set of eigenfunctions.

In contrast to this conventional case, if we want to obtain an invariant and a set of STA eigenfunctions for the Hamiltonian \eqref{ham_ho}, we can express the Hamiltonian in momentum space,
in such a way that the kinetic term $ \partial_z $ becomes multiplicative
\begin{equation} % Hamiltonian in momentum space
	H_0 = \frac{\Omega(t)}{2\chi}p^2 -\hbar^2 \frac{N}{2}\partial_{p^2} \:. \\
	\label{eq:ham_ho_momentum}
\end{equation}

By comparing the aforementioned Hamiltonian and \eqref{eq:HamiltonianHarmonicOscillator}, we can see that the two are equivalent and the following mapping will allow us to write the STA eigenfunctions for the Hamiltonian in Eq.~\eqref{eq:ham_ho_momentum}:
\begin{align} % Mapping between Josephson and harmonic oscillator
	m & \rightarrow \frac{1}{N} \label{eq:mapping_m} \\
	m \omega^2(t) & \rightarrow 	\frac{\Omega(t)}{\chi} \label{eq:mapping_omega}.
\end{align}
We can now write down the instantaneous eigenstate of the Lewis-Riesenfeld invariant of the Hamiltonian \eqref{eq:ham_ho_momentum} as
\begin{eqnarray} % STA eigenstate in momentum space
\tilde\chi_n(p,t) &=&	C^{1/4} \frac{1}{\pi^{1/4}}
\frac{1}{\sqrt{2^n n! b}} \nonumber \\ % Normalisation term
&\times&	\exp\left\{i(2n + 1)\ \int_{0}^{\tau}d s ~ \frac{C}{b^{2}(s)}\right\} % time dependent phase factor 
	\nonumber \\ % new line
&\times&    \exp\left\{- \frac{p^{2}}{2} 
		\left(\frac{C}{b^{2}} - \frac{i\dot{b}}{2b}\right)
	\right\}\nonumber \\
&\times&	\mathcal{H}_{n}\left[C^{1/2}\frac{p}{b}\right]
\label{eq:chi_momentum}
\end{eqnarray}
where $C=\sqrt{\frac{\Omega_0 N}{4\chi}} = \frac{N}{2\sqrt{\Lambda_0}} $. 

In order to evaluate the eSTA corrections, it is necessary to evaluate the auxiliary functions $ G_n $ and $\mathbf{K}_n$ [cf. Eqs.~\eqref{eq:Gn} and \eqref{eq:Kn}, respectively].
%Said terms involve an operator $ \Delta H $ (\eqref{eq:DeltaH}) and $ \nabla H $ respectively and 
To this end, it is useful to calculate the Fourier transform of the invariant eigenstates and thereby get these invariant eigenstates in the position respresentation:
\begin{eqnarray} % Fourier transform of chi0
\lefteqn{\chi_n (z,t) = \mathfrak{F} \left[ \tilde{\chi}_n\right] (z,t)} & & \nonumber\\
 &=& 
		C^{1/4} \frac{1}{\pi^{1/4}} \frac{1}{\sqrt{2^n n! b}} \nonumber\\% Normalisation term
&\times& \exp\left\{i(2n + 1)\ \int_{0}^{\tau}d s ~ \frac{C}{b^{2}(s)}\right\} % time dependent phase factor 
	\nonumber \\ % new line
&\times&	  \frac{1}{\sqrt{2\pi h}} \int_{-\infty}^{+\infty} dp \  \exp\left\{ i\frac{p}{\hbar}z  \right\}
	\nonumber \\ % new line
&\times&    \exp\left\{- \frac{p^{2}}{2} 
		\left(\frac{C}{b^{2}} - \frac{i\dot{b}}{2b}\right)
	\right\}
\mathcal{H}_{n}\left(C^{1/2}\frac{p}{b}\right)
\label{eq:chiFourier}
\end{eqnarray}
where the integral acts only on the part of the wave function that depends on the variable $ p $.
Given that the integrand in Eq.~\eqref{eq:chiFourier} is a product of a Gaussian and an 
exponential function, this last integral can be solved analytically. In general, the following relation holds true:
\begin{align} 
	\frac{1}{\sqrt{2 \pi}}\int_{-\infty}^{\infty}dk \exp\left\{-\frac{k^2}{2 \alpha^2}\right\}\mathcal{H}_n(\beta k)e^{ikx} = \nonumber \\
	i^n \alpha (2\alpha^2 \beta ^2-1)^{n/2} % normalisation factor
	\exp\left\{-\frac{\alpha^2x^2}{2} \right\} % Gaussian term 
	\nonumber \\ % new line
	\times \mathcal{H}_n\left(\frac{\alpha^{2} \beta }{\sqrt{2 \alpha^2 \beta^2-1}}~x\right) ~. % Hermite polynomial
	\label{eq:FourierTransformGeneral}
\end{align}
Finally, by making use of this last expression, we can obtain 
the corresponding wave function in terms of position representation Eq.~\eqref{eq:chi} where $k_0 = N/(2\sqrt{C})$ and $\beta=b \dot b/(2C)$.

\end{document}